\documentclass[conference]{IEEEtran}
\usepackage{hyperref}
\usepackage{listings}

\usepackage{graphicx}

\ifCLASSINFOpdf
\else
\fi
\ifCLASSOPTIONcompsoc
  \usepackage[caption=false,font=normalsize,labelfont=sf,textfont=sf]{subfig}
\else
  \usepackage[caption=false,font=footnotesize]{subfig}
\fi

\usepackage{xcolor}

\usepackage{xspace}

\newcommand{\cooley}{\textsc{Cooley}\xspace}

\newcommand{\thetasc}{\textsc{Theta}\xspace}

\begin{document}

\newcommand*{\NODEBUG}{}

\ifdefined\DEBUG
\newcommand{\todo}[1]{\textcolor{blue}{[TODO: #1]}}
\newcommand{\george}[1]{\textcolor{magenta}{[GKT: #1]}}
\newcommand{\cameron}[1]{\textcolor{orange}{[CC: #1]}}
\newcommand{\xiaoyong}[1]{\textcolor{red}{XYJ: #1}}
\newcommand{\francois}[1]{\textcolor{blue}{FT: #1}}
\newcommand{\venkat}[1]{\textcolor{green}{VV: #1}}
\fi

\ifdefined\NODEBUG
\newcommand{\george}[1]{}
\newcommand{\cameron}[1]{}
\newcommand{\xiaoyong}[1]{}
\newcommand{\francois}[1]{}
\newcommand{\venkat}[1]{}
\fi

\newcommand{\tomove}[1]{\textcolor{blue}{[TO MOVE: #1]}}
\newcommand{\todo}[1]{\textcolor{red}{[TODO: #1]}}

%
\title{A Benchmarking Study to Evaluate Apache Spark on Large-Scale Supercomputers}


\author{
    \IEEEauthorblockN{George K. Thiruvathukal\IEEEauthorrefmark{1}\IEEEauthorrefmark{4}, Cameron Christensen\IEEEauthorrefmark{2}, Xiaoyong Jin\IEEEauthorrefmark{3}, Fran\c{c}ois Tessier\IEEEauthorrefmark{5}\IEEEauthorrefmark{4}, Venkatram Vishwanath\IEEEauthorrefmark{4}}    	
    \IEEEauthorblockA{\IEEEauthorrefmark{1}Loyola University Chicago, Computer Science Department, gkt@cs.luc.edu}
    \IEEEauthorblockA{\IEEEauthorrefmark{2}University of Utah, Scientific Computing and Imaging Institute, cam@sci.utah.edu}
    \IEEEauthorblockA{\IEEEauthorrefmark{3}Argonne National Laboratory, Computational Science Division, xjin@anl.gov}
    \IEEEauthorblockA{\IEEEauthorrefmark{4}Argonne National Laboratory, Leadership Computing Facility, gkt@anl.gov, venkat@anl.gov}
    \IEEEauthorblockA{\IEEEauthorrefmark{5}Swiss National Supercomputing Center, ETH Zurich, Lugano, Switzerland, ftessier@cscs.ch}
}


%


\maketitle

\begin{abstract}
As dataset sizes increase, data analysis tasks in high performance computing (HPC) are increasingly dependent on sophisticated dataflows and out-of-core methods for efficient system utilization. In addition, as HPC systems grow, memory access and data sharing are becoming performance bottlenecks. Cloud computing employs a data processing paradigm typically built on a loosely connected group of low-cost computing nodes without relying upon shared storage and/or memory. Apache Spark is a popular engine for large-scale data analysis in the cloud, which we have successfully deployed via job submission scripts on production clusters. 

In this paper, we describe common parallel analysis dataflows for both Message Passing Interface (MPI) and cloud based applications. We developed an effective benchmark to measure the performance characteristics of these tasks using both types of systems, specifically comparing MPI/C-based analyses with Spark. The benchmark is a data processing pipeline representative of a typical analytics framework implemented using map-reduce. In the case of Spark, we also consider whether language plays a role by writing tests using both Python and Scala, a language built on the Java Virtual Machine (JVM). 
We include performance results from two large systems at Argonne National Laboratory including \thetasc, a Cray XC40 supercomputer on which our experiments run with 65,536 cores (1024 nodes with 64 cores each). The results of our experiments are discussed in the context of their applicability to future HPC architectures. Beyond understanding performance, our work demonstrates that technologies such as Spark, while typically aimed at multi-tenant cloud-based environments, show promise for data analysis needs in a traditional clustering/supercomputing environment. 


\end{abstract}



%
\IEEEpeerreviewmaketitle


\section{Introduction}
\label{sec:introduction}




In this paper, we are motivated to understand and articulate the characteristics of cloud-based dataflow processing in the context of high-performance data analysis tasks since large HPC systems are beginning to resemble high performance clouds. As such, being able to schedule and run cloud-based data analysis software allows simulations to perform co- or post-processing with a relatively straightforward computational model (map/reduce) based on functional programming. In order to compare the performance and deployment challenges of such applications with traditional MPI-based solutions, we have developed a data processing benchmark using both MPI and the Apache Spark~\cite{Zaharia:2016:ASU:3013530.2934664} cluster-computing framework (hereafter referred to as \textit{Spark}), that we execute on two supercomputers. We compare performance of this benchmark by measuring strong and weak scaling. For Spark, we also examine performance of two supported languages: Python and Scala (the primary language used to author Spark itself). The purpose of these comparisons is not to make a value judgment about which system is best 
but to demonstrate whether Spark is viable in a supercomputing environment. We envision many scenarios where Spark can be incorporated in mixed MPI/Spark environments (e.g. simulations) where an MPI-based computation could be co-scheduled with a Spark-based application, thereby allowing popular higher-level Python/Java/Scala frameworks to be incorporated into supercomputing applications. To keep our focus narrow, we avoid introducing additional layers, such as data loading and out-of-core analyses, though these will be of great interest in future work. The aim of this work is to assess the out-of-box Spark experience on computational clusters. We therefore made no changes to the Spark runtime itself, and limit customizations to only the key Spark properties necessary to utilize the cloud computing framework on HPC systems.

%
%

We use the term \emph{dataflow} to describe a set of operations for the creation, processing and transformation of a given dataset. This differs from a related computational approach from the 1980s \cite{Arvind:1977:IMD:1067625.806559}, in which dataflow is a fine-grain parallel architecture focused on triggering instructions using tokens. Tokens arrive at instruction nodes, which cause an instruction to \emph{fire}, resulting in tokens that can be passed to subsequent instructions. Nor should the term be confused with the compiler principle of data-flow analysis, which is used to perform various code analyses and optimizations, including but not limited to dead-code elimination, common subexpression elimination, and loop unrolling. 

While our efforts do not consider every type of MPI application, such as tightly-coupled simulations, many data-intensive programs built on MPI follow a typical map-reduce pattern when performing multistage analyses, in which data is first loaded or generated, then one or more computations are performed on this data (possibly transforming it in the process), and finally a global operation is utilized to obtain a result. This multistage dataflow is commonly found in many applications.  Some recent papers points to increased use of map-reduce in large-scale scientific and technical computing to support distributed scientific calculations using a single, parallel map/reduce job that were previously done with separate one-node jobs~\cite{5708835}; streaming data analysis of streaming scientific (e.g. sensor) and business (e.g. financial markets) data~\cite{6493198}; and large-scale data science and machine learning.~\cite{Shanahan:2017:LSD:3041021.3051108}. 

Consider, for example, an image processing application for feature extraction. This program would load the set of images to be analyzed, transform this data by color space, and perform an in-place \textit{map} operation on the pixels to identify one or more features. Finally, the \textit{reduce} operation gathers information about the presence or absence of features across all images.
Our benchmark implements a similar multistage set of operations, abstracting some details while ensuring that the actual workload remains realistic and reproducible in real-world situations. It is described in detail in Section~\ref{sec:benchmarking}.

The Spark dataflow creates a resilient distributed dataset (RDD) that can be persisted in memory and/or on disk for the lifetime of the computation. These RDDs can be quickly transformed into {new} RDDs using the \emph{map} operation (the notion of mapping originates in \emph{functional programming languages}~\cite{McCarthy:1962:LPM:1096473}). An earlier system, Hadoop~\cite{Shvachko:2010:HDF:1913798.1914427}, included the map/reduce framework but focused on storage, which is not absolutely required in every meaningful dataflow (per the cited examples above).  Spark also supports map/reduce but is focused on objects in memory which can be persisted on demand or when available memory is exhausted, which is attractive for out-of-core memory workflows.

In addition to understanding how cloud computing frameworks compare with MPI/C, we wanted to understand whether the specific languages used by the cloud frameworks affect overall performance. The two main choices supported by Spark and likely to be of interest to HPC developers are Python, a dynamic language already popular in computational and data science, and Scala, an object-functional JVM language. We are encouraged by the results we are seeing in both of these approaches, and we will speak more generally to their affordances and constraints in our detailed performance analysis and conclusions.

Lastly, we are also intrigued by cloud computing frameworks for software engineering reasons. Python and Java are used by a large number of data analysts and machine learning researchers and feature many libraries that enable this work.
The availability of cloud computing on HPC systems without making a significant compromise to performance will help enable more of these practitioners to maximally utilize available resources. The experiments presented in this work are among the largest studies, both in terms of number of nodes and total core count, and we believe they can be used to help understand and improve cloud computing performance in large-scale supercomputing systems.

Our work to create general-purpose job submission scripts (see~\ref{sec:running}) allows us to create multiple Spark networks of the appropriate size to best enable optimal performance of application-specific code. We currently utilize 
our own job scheduler, but these scripts can be ported (easily) to other popular job schedulers, enabling Spark applications to run on other HPC systems.

In sum, our paper focuses on the following aspects:

\begin{itemize}
\item Comparison of traditional MPI data processing with cloud-computing frameworks
\item Real scalability using up to 65,536 cores on a leadership-class HPC system 
\item Comparison between Python or Scala Spark versions and MPI/C version
\item Identifying and addressing Spark performance overheads
\item Demonstration of portability between two significantly different leadership class systems
\end{itemize}


\section{Related Work}

Bringing cloud-based frameworks such as Spark to supercomputing frameworks is a subject of increased interest in high-performance parallel/distributed computing environments. We consider related work on benchmarking (the most closely related papers to our study), challenges of Spark development, scientific applications using Spark, tuning-related issues, and numerical libraries used in our study for the Python and Scala versions.

\subsection{Benchmarks}

Chiamov et al~\cite{Chaimov:2016:SSH:2907294.2907310} described their experiences porting Spark to large-scale HPC systems. They observed that I/O is the main bottleneck (using Lustre metadata) on supercomputers for this type of High Performance Data Analytics (HPDA) workload. To mitigate that, they developed a file pooling layer and ran experiments using NVRAM buffers (comparable to the on-node SSDs found on \thetasc, one of our testbeds). Some key takeaways from their work include using local storage for the shuffle stage, which improved scalability in their problem to 10,000 cores.  This work also evaluated a configuration with storage attached closer to compute nodes for I/O acceleration. Our paper is mostly focused on large, generated in-core data sets, but we are also evaluating I/O from the shared disk and experienced similar results.

The authors uncovered several scaling issues on HPC systems, which would likely be worse on lower-performing clusters. Fixing the YARN resource manager and improving the block management in the shuffle block manager will benefit performance. Our experiments have mostly been confined to being careful with persistence schemes, especially with how data are serialized (on the Java side in particular, where you want to avoid serializing complex object hierarchies). Our Scala experiments do take advantage of said serialization, which probably explains some of the overheads we are seeing in our performance charts.

Gittens et al~\cite{DBLP:journals/corr/GittensDRRGKLMC16} done a study comparing MPI/C++ and  Spark Versions. In this work, the authors developed three different parallel versions of matrix factorizations and apply them to TB (terabyte) size data sets. Their testbed is the Cray XC40 with up to 1600 nodes. Initial findings confirm a performance gap between MPI/C++ and Spark---a gap that we also observe in our experiments. The performance gaps are attributable to task-related overheads that would not be present in MPI, which creates and schedules its tasks at compile time in typical computations. Similar to Chiamov et al, serialization can play a major role in Spark, even when data are persisted to RAM. Our study is a bit more focused on strong/weak scaling in the presence of these overheads. 

Ringenburg~\cite{CUG16:Ringenburg} considers performance characteristics of HPDA workloads on a Cray XC40 system. The authors analyze the log and workload of two publicly available benchmarks: Intel's HiBench 4.0 Suite and a CX matrix decomposition algorithm. The CX matrix decomposition experiments use up to 960 cores. The presentation provides guidelines for improving performance on this particular Cray machine, including tuning of various Java Virtual Machine parameters, e.g. garbage collection threads, etc. The platform was Spark 1.5 with no local storage available. Our results are based on relatively recent Spark releases (discussed in experimental setup, section~\ref{sec:benchmark-setup}).

Marcu et al~\cite{marcu:hal-01347638} et al and Garcia et al~\cite{García-Gil2017} propose a comparison between Spark and Flink. Marcu et al introduce a methodology based on a set of benchmarks (word count, grep, tera sort, K-means, page rank, connected components, many of which are example programs in the Spark distribution) ran on up to 100 nodes to understand performance in this type of frameworks. Garcia et al shows the results of three ML algorithms running on 10 nodes, 16 cores/node. Our benchmark, while synthetic in nature, runs at much higher node and cores per node counts.

Chunduri et al~\cite{Chunduri:2017:RVX:3126908.3126926} discuss run-to-run variability on \thetasc (one of the supercomputers we performed our study in this paper) a Cray XC40 system, where the authors observed that the MPI\_AllReduce on small message sizes appears to have the highest variability due to inter-job contention. We have also observed about 15\% variability in our experiments. In this paper, we focus on the general scaling behavior, and our Spark benchmarks indicate other overheads playing an even more dominant role, especially when it comes to TCP connections (see Section~\ref{sec:tuning-spark}), we leave the quantification of variability in our future study.

\subsection{Challenges of Spark Deployment} 

Armbrust et al \cite{Armbrust:2015:SSR:2824032.2824080} present feedback from a company deploying Spark to various organizations. This paper presents some of the main difficulties encountered such as large-scale I/O or memory management. The authors developed some memory management features as well as a custom network module. To make Spark more accessible to non-experts, they also wrote an API based on data frames (like in Python or R). Our paper focuses on the Spark RDD, which is the underlying fabric used to implement data frames and data sets. 

Tous et al~\cite{7363768} discuss an optimized deployment of Spark on MareNostrum, the BSC's supercomputer. The authors developed a framework to automate the usage of Spark. They also provide guidelines on using Spark on this system (number of workers, size of the workers). As part of our work, we developed generalized job submission scripts, discussed in detail in Section~\ref{sec:spark_setup}, which allow the Spark daemons to be launched in a generalized way (on two of our clusters), followed by launching application-specific code in Spark (which acts as a container).

\subsection{Scientific Applications}

Yan et al~\cite{7363985} explored the scalability of  Spark on a set of seismic data processing algorithms. To do so, the authors propose to change the way data is given to Spark (templates for seismic data sets).

Souza et al~\cite{souza:lirmm-01620161} present a scalability analysis of Spark through a synthetic implementation of a scientific workflow based on a real use-case in oil and gas domains. They processed 330 GB of data on a 936-cores HPC cluster. The application is not written for Spark. Instead, they used a "black-box" approach (run external program with Spark). One of the outcomes is that the task duration has a better scalability than the number of tasks (typical strong versus weak scaling).

\subsection{Java and Big Data Processing Tools on HPC Systems}
\label{sec:tuning-spark}

Nowicki et al~\cite{10.1007/978-3-319-78054-2_27} present a library for Java called PCJ, whose purpose is to easily allow parallel computation. This library is evaluated on a KNL-based platform and the authors come to the conclusion that Java can run successfully on architectures designed for parallelism. In~\cite{8030575}, the authors also targets the Intel KNL architecture and describe how they evaluated Hadoop on it through a custom plugin. Again, one of the conclusions is that it's feasible to leverage intra-node parallelism with Big Data processing frameworks.

Jacobsen et al~\cite{CUG16:Jacobsen} describes the SLURM implementation on large-scale Cray systems, including a KNL system similar to \thetasc. In particular, the team found the need for Linux kernel tuning to increase max TCP connections to address SYN (connection) backlog. As Spark itself relies on a significant number of connections from executor nodes to the driver, we see some evidence of connection issues in our own experiments, resulting in significant overheads. We have not done these Linux kernel level tunings in our production cluster but hope to address the issue in future work.

\section{About Apache Spark}
\label{sec:about_spark}

Spark is a general purpose cluster computing system similar to Hadoop. It provides a new data abstraction that facilitates fast sharing and history-based resilience, as well as an expanded set of data transformations and actions, in addition to traditional map-reduce. In addition, Spark provides a streaming processing abstraction as well as bindings to common processing languages such as GraphX, R, and MLlib.


When a dataset is loaded by Spark, it becomes an immutable Resilient Distributed Data (RDD) collection. This abstraction allows the data to be treated as a whole when in fact it may
be partitioned across many nodes of a distributed system. Each partition also contains the history of transformations with which it was created,
called a \emph{lineage}, with which the partition can be recomputed if necessary, such as in the case of a node failure. This lineage is a more compact form of resiliency compared to data duplication as utilized by Hadoop.

Spark is \emph{lazy}, and this philosophy underlies much of its design. Computations will not be performed until their result is requested and data will not be consolidated or repartitioned unless explicitly requested. 
In order to facilitate flexible and generalizable task parallelism RDDs are read-only data structures and therefore lazy evaluation can result in faster overall run times since it avoids unnecessary memory allocation. In contrast, MPI/C computations can be carried out in place and therefore the notion of deferring a computation is not relevant. 

An RDD is split into \emph{partitions} whose size is at minimum the size of a block on whatever storage device is being utilized. Each partition is further divided into \emph{records}, typically a
single line for text processing, or an entire binary file for binary data. Binary data records can be explicitly specified. Large binary files will be broken down into multiple partitions only if these
partitions can themselves be divided into records.

Spark distributes the blocks/data among the workers, if it does at all.  Spark supports fault tolerance among the workers. For example, if a worker is lost, processing can continue. Although node failure rates are expected to be low in small-to-medium size clusters, larger clusters are more likely to see single node failures, thus making Spark potentially compelling even in highly-reliable supercomputing clusters. As is well-known in the MPI community, attaining such functionality requires application-specific checkpointing and/or a fault-tolerant runtime, both of which incur significant overheads. With Spark, the transparent support for fault tolerance allows the application code to be written without such overheads (at least in theory). While not the subject of this paper, we are intrigued by the potential to look at performance in future work, especially in the presence of one or two node failures.


RDDs can be \emph{persisted} or \emph{cached} in memory, on disk, a combination of the two, or off heap. In our experiments, we focus on RDDs that persist to RAM. We also have designed our experiments to consider out-of-core persistence strategies and spill rates (which we can simulate by multiplying the number of blocks, the number of partitions, or both). Partitions that do not fit in memory will be recomputed from their history when they are needed. This recomputation is possible because RDDs, which include the data \emph{lineage}, are basically Scala immutable collections and are manipulated using map (to create a new RDD) and reduce (to gather results from nodes and apply a function between all pairs). In our benchmark, we are primarily considering mapping dataflow performance but also report initial results on reduce performance.






There are many ways to launch Spark. In cloud environments, Mesos or YARN are popular cluster managers. Spark, however, also provides a built-in standalone deploy mode, which is achieved with a collection of start/stop scripts. These are easy to deploy in our job scheduling environment, where a set of nodes is allocated to the user for a time duration requested at job submission. When our job starts, we launch Spark in the standalone deploy mode, with one Spark master on one node and one Spark worker on each node of our allocation, and submit our benchmark script to the Spark master.








\subsection{Submitting Spark Jobs on our Clusters}
\label{sec:running}


The supercomputers we use for this paper schedule and execute compute jobs via a job scheduler.
A user can request a specific number of compute nodes and a fixed maximum wall clock time for a compute job given as a binary or a script.
Such job will then wait in the queue and get launched when the requested compute resources become available.
When the job starts, the user, in a job script, typically launches the executable through computing resource specific methods, such as \verb|mpirun| on \cooley or \verb|aprun| on \thetasc. We will discuss our computing resources in detail in Section~\ref{sec:benchmark-machines}.

In collaboration with Argonne National Laboratory, We have developed a set of scripts that automate the job submission procedure, where the main user facing interface is a bash script.

\subsection{Starting the Spark Framework}
\label{sec:spark_setup}

In order to run Spark jobs on our two HPC system we need to first start up the framework. This entails specification of the size of the desired cloud-computing cluster as well as the internal URL from which it can be accessed. Since it is designed to work with a general set of nodes accessible via SSH, we use the built-in Spark standalone cluster mode in our study. All of this is automated in our scripts.

On our first test-bed, we find that the Spark built-in scripts, which uses SSH, work well by default, once we overwrite the SSH command in a bash function and provide essential environment variables through SSH invocation.






On the second test-bed, due to the large amount of nodes, we use the Cray recommended command \verb|aprun| to launch the Spark master and workers.

\subsection{Running the Spark Job}

Once Spark is started in cluster mode, we now have the equivalent of a local cloud, albeit ephemeral, that we can use to run one or more Spark jobs. 
We use the environment variable passed through our modified SSH command or the \verb|aprun| to specify the Spark master URI, and use the Spark built-in \verb|bin/spark-submit| on the Spark master node to submit a Python or Java/Scala job for execution on the cluster.
These are also automated in our scripts.


\section{Benchmarking}
\label{sec:benchmarking}

We've developed comprehensive benchmark tests for each language (MPI/C, Python, and Scala) designed to work in parallel on large, distributed array data structures. From the user's point of view these data structures are continuous, but on the system they are distributed across all available nodes as blocks, and the size of each block can be explicitly specified. Smaller blocks have the advantage of being able to be rapidly transferred between nodes while larger blocks require less overhead and can therefore be processed more efficiently. The data can be loaded from disk, from the network, or generated directly on the node itself by the application.

Our benchmarks are designed to emulate a typical image analysis dataflow based on the \textit{map-reduce} paradigm, which consists of four stages: data loading/generation, then some number of operations executed in parallel on that data, and finally a global reduction to obtain a final result. Each stage can be timed separately in order to more precisely analyze overall execution time.


The following parameters apply to all of the benchmarks:

\begin{itemize}
\item blocks: Number of blocks to be created
\item block size: Basic block size
\item nodes: Number of nodes actually present in the cluster
\item cores: Number of cores per node in the cluster
\end{itemize}


\subsection{Benchmark Machines}
\label{sec:benchmark-machines}



\subsubsection{\cooley}

Our first test-bed is a mid-size cluster at the Argonne Leadership Computing Facility (ALCF), aimed primarily but not exclusively at support high-performance visualization processing and analytics. It has a total of 126 compute nodes. Each node has two 2.4 GHz Intel Haswell E5-2620 v3 processors (6 cores per CPU, 12 cores total), with 384GB RAM. This system uses FDR Infiniband interconnect.



\subsubsection{\thetasc}

The second test-bed is a large-scale Cray XC40 supercomputer with only a single compute node type: the Intel Knights Landing 7230 processors. It features 16 GiB of MCDRAM and 192 GiB of DDR4 per node. Each node has one 1.3 GHz Intel Xeon Phi 7230 processor with 64 cores, and each core has 4 simultaneous multhreading (SMT) hardware threads available. At the time of writing, the platform has a total of 4392 compute nodes (281,088 cores). For our experiments, we were able to use 1024 of these nodes. The Cray Aries network is the high-speed interconnect used on this system. This network is a 3-level Dragonfly topology.



\subsection{Benchmark Setup}
\label{sec:benchmark-setup}

We study the strong and weak scaling behavior of our three code bases on \cooley and \thetasc using simple, but still realistic, parameters for our tests. Though we have developed a comprehensive benchmark, in this work we only compare performance of data {generated} by a simulation or other analysis, not data loaded from disk. Another affordance of  Spark is its innate support for of out-of-core execution \emph{implicitly} and \emph{by default} using various strategies to cache to RAM, to disk, and even combinations of the two. This is something that is simply not possible in traditional ahead-of-time compiled languages such as C without essentially reproducing many of the ideas of the JVM, building a memory-management framework, or using memory caching services such as memcached. While our benchmark has been designed to allow for \emph{data spilling} by adjusting any of the benchmark parameters (blocks, block size, or the partition multiplier), the tests presented in this work deliberately avoid this in order to keep the focus narrowed on the difference between cloud computing and MPI/C without the additional layers required to facilitate out-of-core memory management.

All of our Spark benchmarks are performed with the latest release of Apache Spark version 2.3.2. As we focus on the out-of-box experience of Spark usage on supercomputers, we use the Spark binary package pre-built for Apache Hadoop 2.7 and later, directly downloaded from Spark website. On both systems we use Scala version 2.11.8.

On \cooley, we used the server Java SE version \texttt{1.8.0\_60} and Python version \texttt{3.5.1} using Anaconda 4.0.0 (64-bit), and the Intel C Compiler (icc) version \texttt{18.0.3}.
On \thetasc, we used the Java SE version \texttt{1.8.0\_51}. Intel Distribution for Python version \texttt{3.5.2} and the Intel C compiler (icc) \texttt{18.0.0}.

Our benchmark makes use of NumPy~\cite{Walt:2011:NAS:1957373.1957466}, which is one of the most commonly used linear algebra/numerical analysis libraries in Python. This library allows us to work with large-dimension array structures and perform the benchmark operation on dense vectors as efficiently as possible in an interpreted language. The NumPy design has also been making its way to other frameworks, including the Breeze Scala library~\cite{BreezeScala} that we also used in our Scala version of the performance benchmark.

Both NumPy and Breeze are needed, simply put, because the native array support in Python and Scala lacks the ability to work with dense vector and matrix structures, especially when it comes to supporting higher order operations associated with linear algebra.

On the other hand, for our MPI/C implementation, we use a straightforward loop with in-place operations for the linear algebra.  As the actual computation performed in the benchmark is mostly limited by the memory bandwidth, we expect this setup accurately reflects the typical scaling behavior.

We perform benchmark on \cooley using 1 to 96 nodes, while using 128 to 1,024 nodes on \thetasc. In the strong scaling study, we fix the number of total blocks to be 9,216 on \cooley, and 131,072 on \thetasc. In the weak scaling study, we fix the number of blocks per node to be 768 on \cooley, and 512 on \thetasc. In both cases, the block size is kept to be one, meaning $2^{20}$ vectors, each with three IEEE double precision floating point elements. The C/MPI implementation uses 12 ranks per \cooley node, and 128 ranks per \thetasc node. Both Scala and Python implementations partition their RDD in 12 parts per \cooley node, and 256 parts per \thetasc node.

\subsection{Spark Parameters}
\label{sec:spark_params}

Since we are executing tests that involve a much larger number of nodes and cores than those conducted on \cooley, we had to disable the Spark \emph{heartbeat} because it overwhelmed the network with hundreds of TCP/IP connections per second and significantly disturbed performance.
See Section~\ref{sec:tuning-spark} for an explanation of potential TCP/IP issues others have experienced in large-scale Cray systems similar to ours. 
We have also significantly increased Spark parameters related the network timeout, in order to eliminate the false error reports of network timeout, where in fact the expensive network RPC is flooding the Linux TCP/IP stack due to the extreme high node count.

We have increased the Spark driver memory and executor memory, according to our machine resources.  We also enabled parallel garbage collector (GC) in JVM with limited amount of GC threads, 2 on \cooley and 8 on \thetasc.


\begin{figure}[tb] 
    \includegraphics[width=\linewidth]{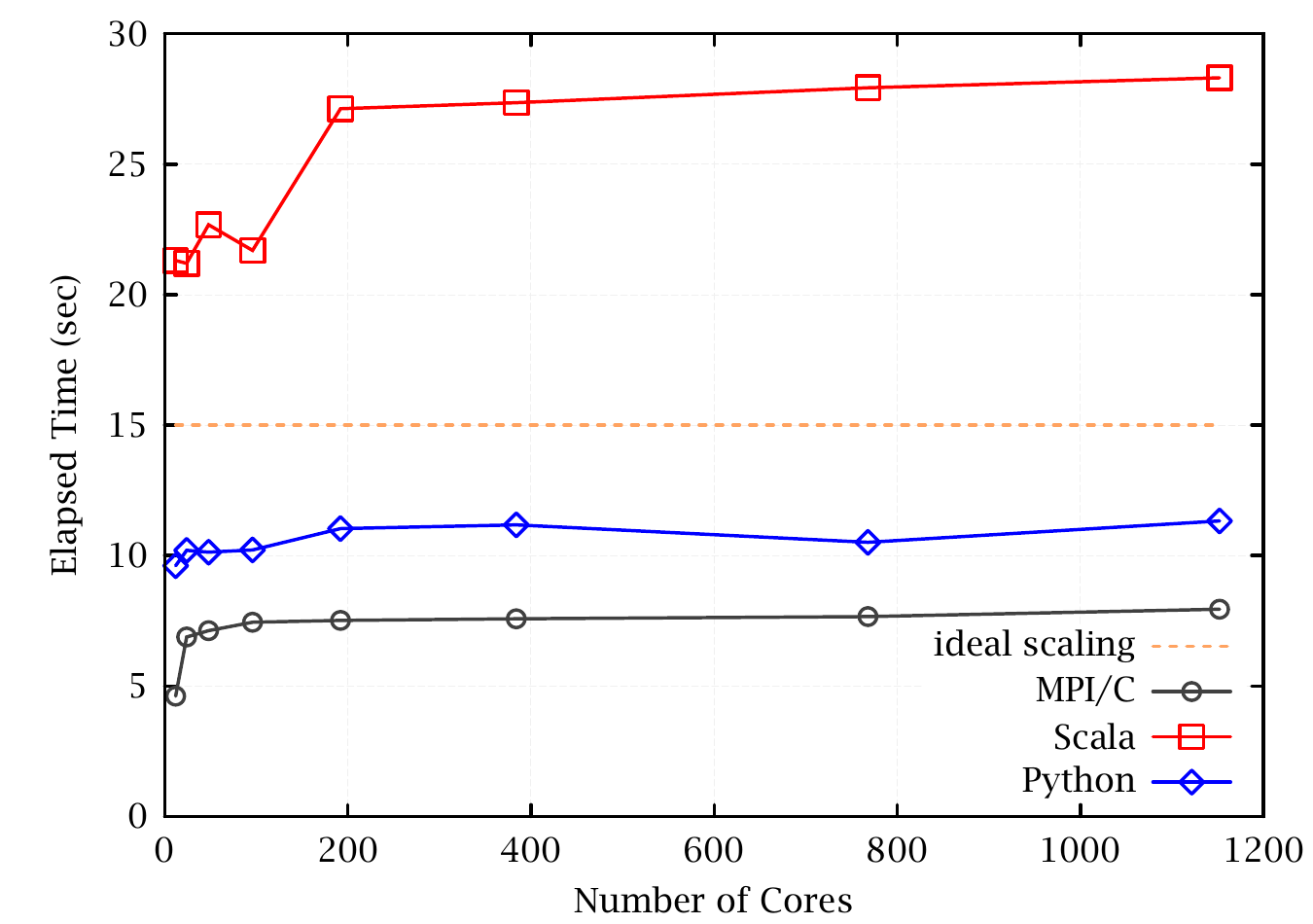}
    \caption{Weak scaling results for \cooley.}
    \label{fig:cooley-weak-scaling} 
\end{figure}

\begin{figure}[tb] 
    \includegraphics[width=\linewidth]{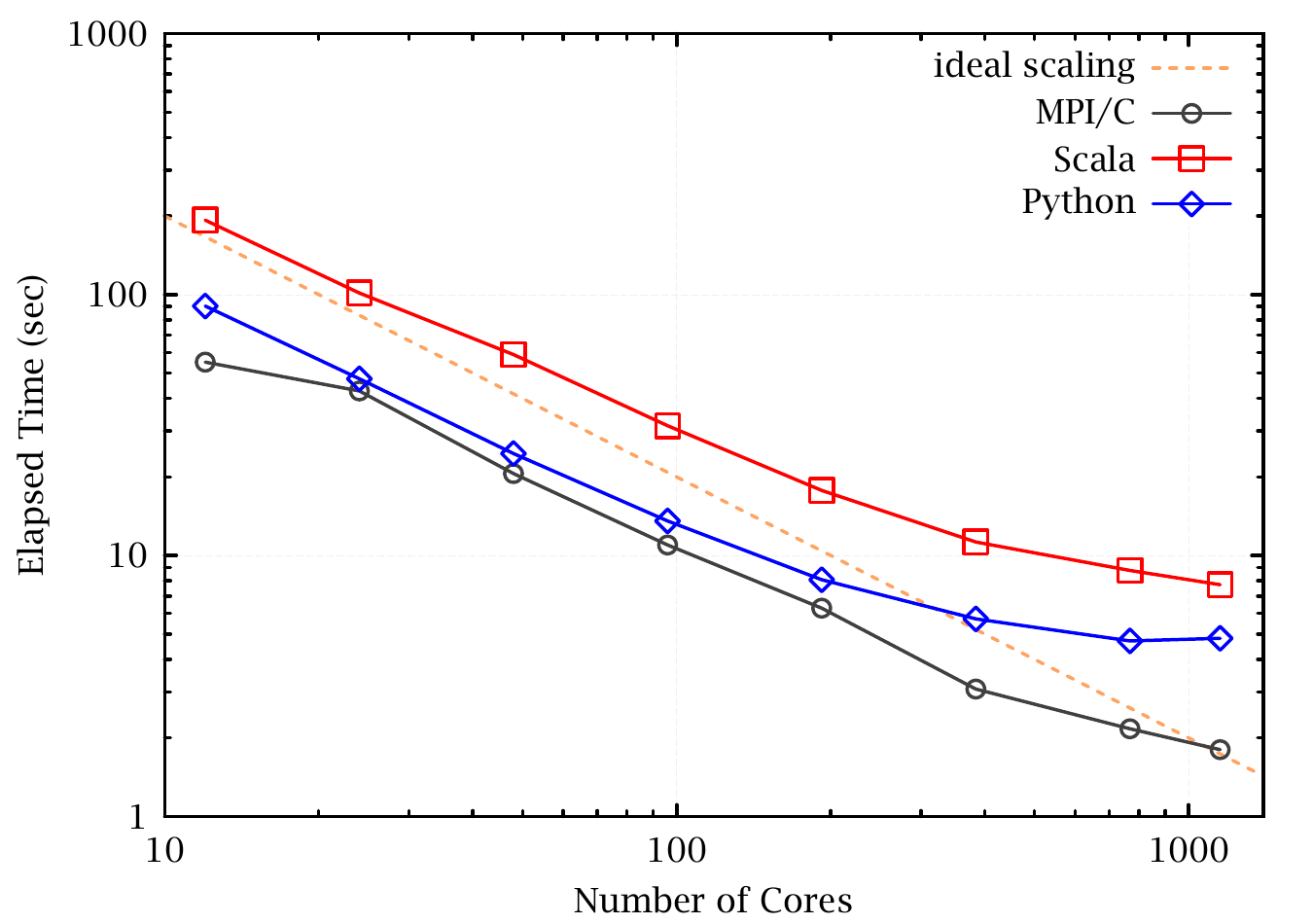}
  \caption{Strong scaling results for \cooley (log scale).}
  \label{fig:strong-scaling-cooley} 
\end{figure}

\begin{figure}[tb]
  \includegraphics[width=\linewidth]{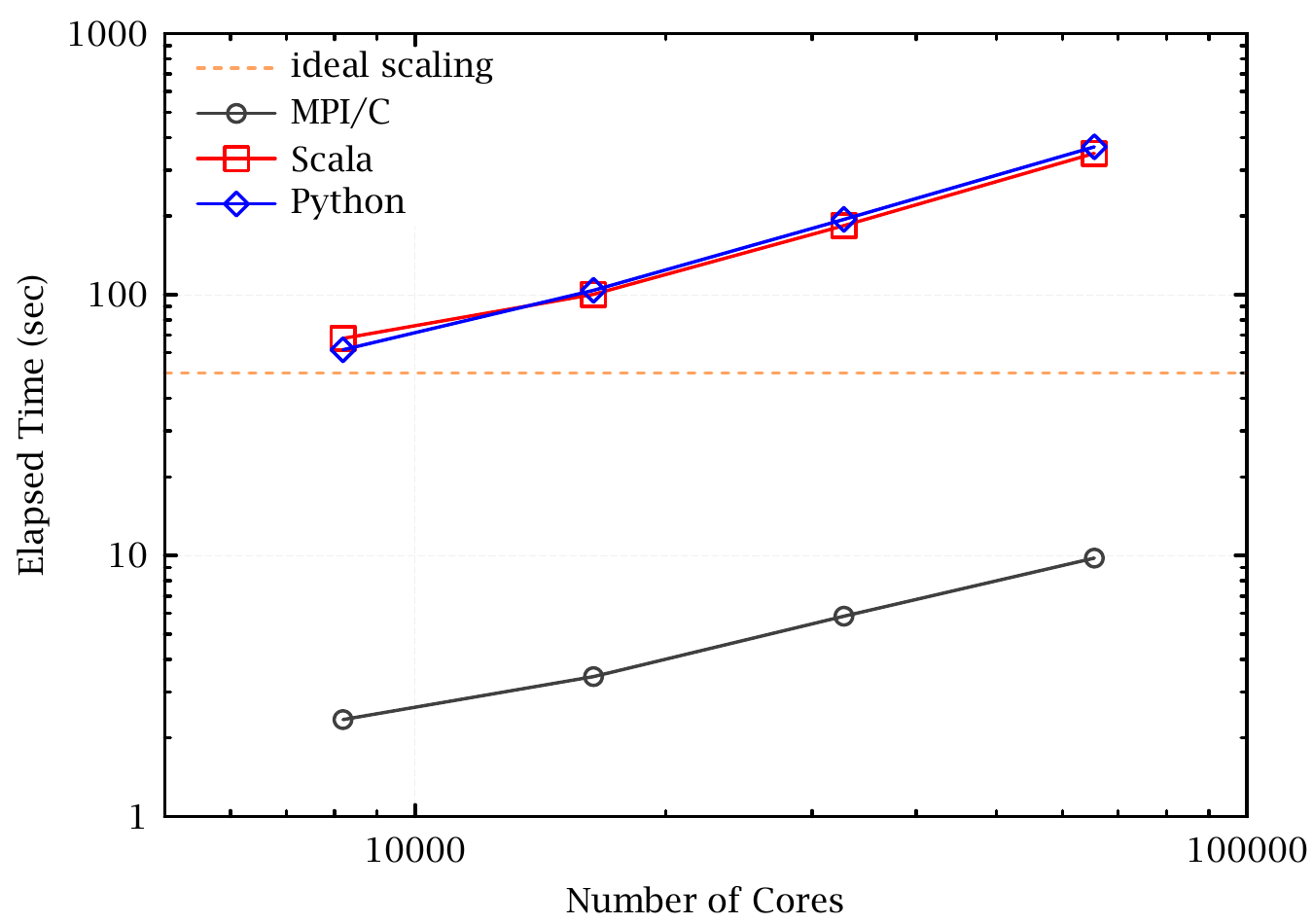}
  \caption{Weak scaling results for \thetasc (log scale).}
  \label{fig:theta-weak-scaling} 
\end{figure}

\begin{figure}[tb] 
   \includegraphics[width=\linewidth]{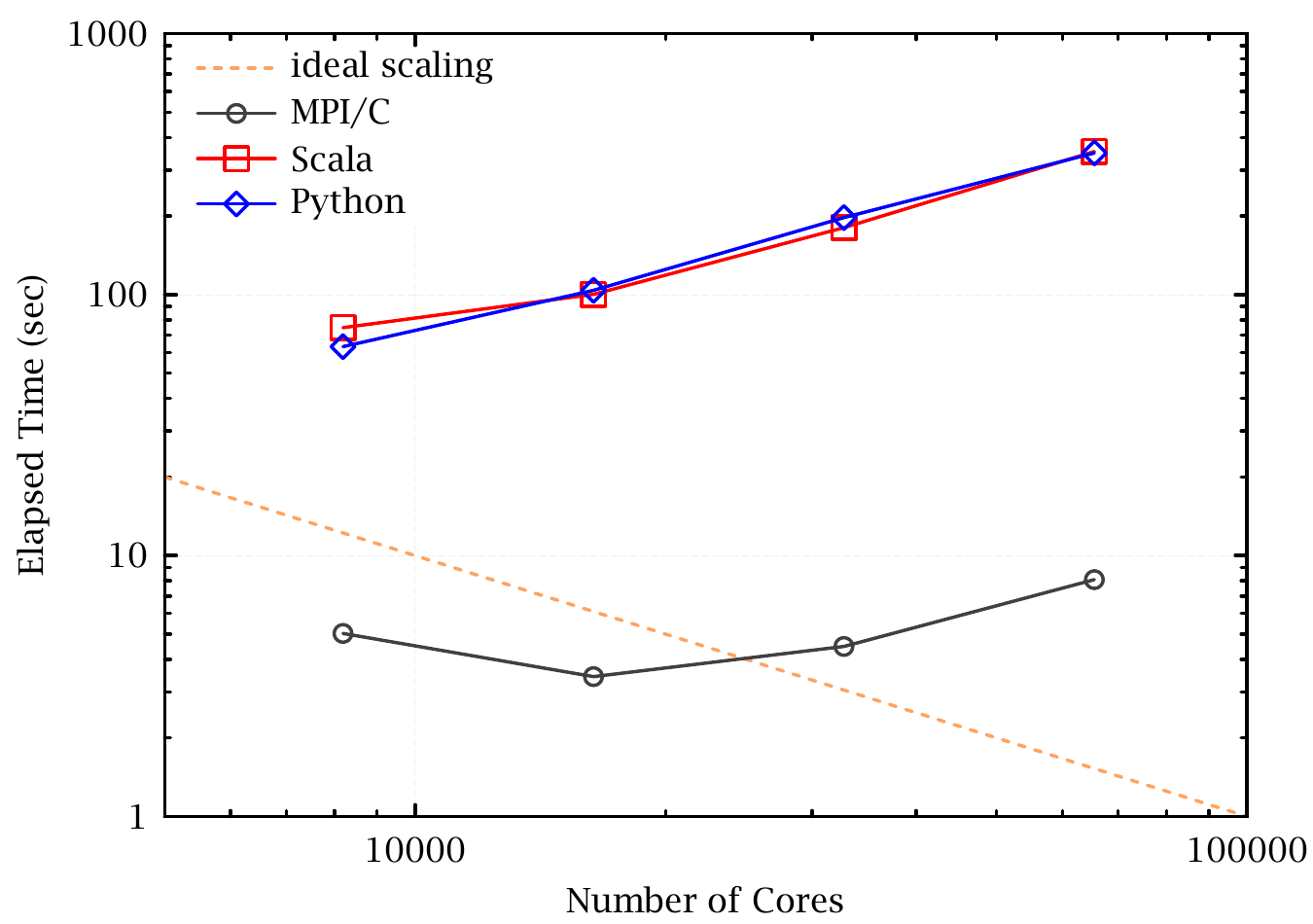}
  \caption{Strong scaling results for \thetasc (log scale).}
  \label{fig:theta-strong-scaling} 
\end{figure}

\section{Results}

Now we present the results of our large benchmark runs on \cooley and \thetasc. On \cooley we used up to 1,152 cores (96 12-core nodes), and on \thetasc we executed much larger runs, using up to 65,536 cores (1024 64-core nodes).
Our experiments were conducted for both strong and weak scaling on each system. Weak scaling, the simplest type of performance comparison, is when the size of the dataset to be analyzed grows proportionally to the size of the system used for its analysis. Ideal weak scaling should be level and independent of the number of cores. Strong scaling attempts to solve a large problem more quickly by keeping the input data size constant while increasing the resources used for its analysis. Ideal strong scaling should be inversely proportional to the number of cores. 
 
In addition, we performed a few tests on each system, probing the throughput and reliability of the interconnect.
We expect Spark is prone to network issues and thread contention.
In contrast to the execution model of MPI, the standalone scheduler in Spark requires message passing in order to actively manage jobs among available workers.

Our benchmark results presented compare dataflow execution times for MPI/C, Spark Python, and Spark Scala. For each of the tests, we provide a figure comparing overall performance. We also show the trend of an ideal scaling (not based on a particular value).

\subsection{\cooley and \thetasc Networking Overheads}


The MPI libraries on both \cooley and \thetasc use API that directly uses the interconnect fabric, the InfiniBand and the Aries, respectively.
Apache Spark communicates at the TCP layer over virtual Ethernet adapters that runs on top of interconnect fabric.  This negatively affects the communication performance of Apache Spark, because all the traffic is going through the normal TCP/IP stack, which requires system calls, memory copies, and network protocols, to run on the CPU, in contrast to MPI that sends traffic dirctly to the interconnect hardware.  Similar observations can be found in reference~\cite{CUG16:Jacobsen}.

To investigate further, we tested point-to-point throughput over the network between two nodes using the \texttt{iperf3}~\cite{iperf3} application.
On \cooley, the \texttt{iperf3} reaches maximum about 16 Gbits/sec using a single connection, which is nearly one third of the MPI point-to-point bandwidth at about 6 GB/sec.
On \thetasc, the \texttt{iperf3} reaches maximum about 14 Gbits/sec using a single connection, which is less than one fifth of the MPI point-to-point bandwidth at about 8 GB/sec.

As Spark uses the open-source Netty~\cite{netty} communication library framework, which uses the Java New IO (NIO) libraries, we wrote a simple benchmarking code using Netty directly to check the network reliability.

Our Netty benchmark shows that the communication on \thetasc is consistently worse than \cooley.
Establishing 8192 connections on \cooley takes 0.5 seconds, while on \thetasc it takes 2 to 3 seconds.
These overheads will affect Apache Spark, simply because its architecture requires a large number of connections to be made to the driver (master).
The maximum response time among these connections is about $20\pm 5$ milliseconds on \cooley,
and $45\pm 10$ milliseconds on \thetasc.
In addition, on \thetasc, among 16 different runs, there are 5 cases where 1 connection failed out of 8192 connections.

The intermittent connection failures, on the one hand, reinforce the use case of Spark, which has builtin fault tolerance and was able to complete all runs successfully in the presence of failures (which only occurred at high node counts).
On the other hand, it increases the run time of Apache Spark.

In addition to the network, each CPU core on \thetasc has lower clock frequency than the CPU cores on \cooley, 1.3 GHz versus 2.4 GHz, respectively, which contributes to poor TCP performance on \thetasc, as the data sending over the TCP/IP stack requires CPU.
Spark potentially has additional performance overhead, because the lower single core performance could worsen thread contention for the Spark scheduler. 
The combination of these issues (unoptimized TCP/IP, point-to-point bandwidth, and connection failures at high node counts) explains the overheads we see in our results.
Nevertheless, we are encouraged that once the Spark network is bootstrapped, the results are encouraging.





\subsection{\cooley}

The benchmarks conducted on \cooley, a middle-tier supercomputer, used up to 1,152 cores (96 12-core nodes) and required no customization of the Spark framework. For both weak and strong scaling, the results of utilizing cloud computing are highly comparable to those of traditional MPI/C. This suggests the use of such frameworks could be valuable and immediately useful for some research.

\paragraph{Weak Scaling}


Figure~\ref{fig:cooley-weak-scaling} shows the overall weak scaling comparison of MPI/C with Spark-based Python and Spark-based Scala tests. We can notice that the Python version is about 50\% slower than the MPI/C version, while the Scala version is about 3X slower. The overall time is mostly flat, indicating a good weak scaling from 12 cores up to 1,152 cores. While the difference between MPI/C and Spark is explained in the previous subsection, the performance gap between the Scala and the Python versions is more surprising. We postulate that the extra overhead in the PySpark serialization and RDD context switches could be the potential cause of the reversed performance, and we will investigate this behavior further in our future study. 


\paragraph{Strong Scaling}


Figure~\ref{fig:strong-scaling-cooley} shows our strong scaling experiment comparing MPI/C with Spark-based Python and Spark-based Scala tests on the same platform. 
The figures utilize a logarithmic scale to more clearly show the time elapsed since the timings become faster inversely proportional to the core counts for this set of tests. We observed that the speed of the Python implementation approaches the MPI/C version from around 24 to 192 cores, while the Scala version stays about 2X the Python version. This result is correlated with our observations on weak scaling experiments. While the MPI/C scales extremely well, the Spark implementations, especially the Python code, start to show some level of departure from ideal strong scaling above 384 cores. Nevertheless, those results are still reasonable and show how a data processing tool can take advantage of a HPC system.



\subsection{\thetasc}

Next we present results from our benchmark runs on \thetasc, a much larger system than \cooley. On \thetasc we start our scaling study with 8192 cores (128 64-core nodes), the smallest size allocation for the default queue of the job manager on this system. The following results show a more significant difference in overall performance for which we suspect network overhead may be responsible with the increased parallelization, an indication we observe on \cooley jobs using more than 768 cores. Another suggestion of these performance differences being caused by networking issues was the necessity to disable the Spark \textit{heartbeat} which otherwise overwhelmed the system as mentioned in~\ref{sec:spark_params}.


\paragraph{Weak Scaling}

Figure~\ref{fig:theta-weak-scaling} shows the overall weak scaling comparison of MPI/C with Spark-based Python and Spark-based Scala tests. The overall weak scaling performance with increasing node counts seems to be similar in how it scales for both Spark and MPI/C, though the Spark runs are about 30X longer than MPI/C. The figure utilizes a logarithmic scale for the axes to depict the similar scaling behaviors: instead of the ideal scaling, which should be flat, the timing clearly increases linearly for all three benchmarks. We expect these networking issues can be resolved. At the moment it means that larger systems such as the one we use here might require further tuning (e.g. TCP/IP and possibly further JVM tuning) to offer a satisfactory scalability.



\paragraph{Strong Scaling}

The strong scaling experiments carried out on \thetasc are depicted in Figure~\ref{fig:theta-strong-scaling}. Again, the figure utilizes a logarithmic scale as in the weak scaling figure above and shows similarly poor scaling for the three benchmarks, all of which depart from ideal scaling, which should be inversely proportional to the number of cores. Similarly to the weak scaling tests, the Spark-based runs are around 30X slower than the MPI/C-based runs. Unlike experiments on \cooley, we can notice here that both versions of our Spark benchmark performs the same way. 



\section{Source Code}
\label{sec:source-code}

The source code for all of our work can be downloaded from GitHub. The Python, Scala, and MPI benchmark codes described in this paper can be found in our GitHub organization, \url{https://github.com/SparkHPC/}, under the repositories, \verb|simplemap-spark-python|, \verb|simplemap-spark-scala|, and \verb|simplemap-mpi-c|. This GitHub organization also contains the framework startup scripts, and job launching scripts described in Section~\ref{sec:spark_setup} and \ref{sec:running}, under the repository, \verb|Spark_Job|. We welcome contributions that enable this framework to be used with other machines. Our run scripts for both machines and results are under the repository, \verb|spark-benchmark-study|.
This GitHub organization page also contains links to related work, including demonstrations of how to use Apache Spark with Jupyter Python notebooks (other work in progress by our team).

\section{Discussion and Future Work}

This work is the beginning of a more detailed understanding of the performance of Apache Spark for HPC dataflows, which have a longstanding tradition of being done in MPI C/C++. Spark dataflows can employ linear algebra libraries like NumPy (for Python) and Breeze (for Scala) aimed at providing near-native C performance. Related work has already shown that MPI C/C++ continues to perform better than variuos alternatives, which therefore helped to focus our efforts on understanding what is possible in Python and Scala (and Java by association). Executing Spark computations on thousands of nodes using tens of thousands of cores may provide a plausible addition to the currently available tools on these systems. And overheads aside, operating at scale on supercomputers makes cloud-based frameworks a viable approach for co-processing and/or postprocessing of more optimized ``native'' computations, such as simulations, etc. Although overheads associated with Java and Apache Spark are significant, the results show that scaling to a large number of nodes and core counts is not only possible but a promising direction.


We have been successful to overcome many challenges to get Spark running on leadership class supercomputers, e.g. \thetasc. With an eye to future leadership systems, work remains to be done, especially when it comes to ameliorating the effects of Spark overheads and ensuring full use of HPC architectures.
Alleviating these overheads will require vendor commitments to tune Java for large core counts and Apache Spark for a large number of connections to the master node.
The code, scripts, and results we've shared will be of value to researchers who want to evaluate the efficacy of cloud-based frameworks on new supercomputing systems.


\section*{Acknowledgment}

The submitted manuscript has been created by UChicago Argonne, LLC, Operator of Argonne National Laboratory (``Argonne"). Argonne, a U.S. Department of Energy Office of Science laboratory, is operated under Contract No. DE-AC02-06CH11357. The U.S. Government retains for itself, and others acting on its behalf, a paid-up nonexclusive, irrevocable worldwide license in said article to reproduce, prepare derivative works, distribute copies to the public, and perform publicly and display publicly, by or on behalf of the Government.

Special thanks to Professor Valerio Pascucci for providing valuable feedback and suggestions for improving this work.



\bibliographystyle{IEEEtran}
%

\bibliography{refs}

\end{document}